\newcommand{\op}[1]{\hat{#1}}
\newcommand{\openone}{\leavevmode\hbox{\small1\normalsize\kern-.33em1}}

\documentclass[10pt]{iopart}
\usepackage{iopams,bm,cite,graphicx,mathptmx,url,color}

\eqnobysec

\begin{document}

\title{Complementarity and phases in ${SU}(3)$}

\author{H. de Guise$^1$, A. Vourdas$^{2}$ and L. L.
  S\'{a}nchez-Soto$^{3}$}

\address{$^1$ Department of Physics, Lakehead University, Thunder Bay,
  ON P7B 5E1, Canada}

\address{$^2$ Department of Computing, University of Bradford,
  Bradford BD7 1DP, United Kingdom}

\address{$^3$ Departamento de \'Optica, Facultad de F\'{\i}sica,
  Universidad Complutense, 28040~Madrid, Spain}

\date{\today}

\begin{abstract}
  Phase operators and phase states are introduced for irreducible
  representations of the Lie algebra $\mathfrak{su}(3)$ using a polar
  decomposition of ladder operators.  In contradistinction with
  $\mathfrak{su}(2)$, it is found that the $\mathfrak{su}(3)$ polar
  decomposition does not uniquely determine a Hermitian phase
  operator. We describe two possible ways of proceeding: one based in
  imposing $SU(2)$ invariance and the other based on the idea of
  complementarity. The generalization of these results to $SU(n)$ is
  sketched.
\end{abstract}

\section{Introduction}

Phase is a unique concept for the proper understanding of classical
optical phenomena.  It is therefore surprising that, at a foundational
level, quantum optics can apparently subsist without a quantum
phase. One can use, for example, the better behaved field-quadrature
operators~\cite{Scully:1997fk}, or be content with a pragmatic
approach in which phase is a parameter that can be efficiently
estimated~\cite{hra95,hb93a,Giovannetti:2004fj,Pezze:2008yq,
  Bahder:2011fr,Dem:2011rt}.  It is equally possible to represent
states as quasidistribution functions in phase space and specify their
phase properties by classical
angles~\cite{sbp92,lp93b,fvs93a,Fiurasek:2000mz,Dubin:2000uq,Schleich:2001}.
Finally, it is also entirely reasonable to approach the problem from
an operational perspective~\cite{sw84,ssw89,nfm91,nfm92,nfm92a} that
emphasizes the apparatus involved in measurements on the system; the
phase then refers to a feature of this apparatus.

However, if one adheres to the orthodox interpretation of quantum
mechanics and regards phase as a physical property then surely it
ought to be represented by a Hermitian operator. In other words, the
phase variable should be subject to quantization and, for a
sufficiently small number of particles, quantized phase effects should
be accessible to experimentation.

Despite the difficulties borne out of the first and eminent
attempts~\cite{lon26,dir27,hei54}, and ultimately ascribed to the
semiboundedness of the eigenvalue spectrum of the number operator,
significant progress has been achieved in the last fews years in
clarifying the status of a quantum phase operator.  The primary
objective in this context has been the description of the phase of a
single-mode field, or, equivalently, of a harmonic oscillator. The
progress made is manifest and the work on this subject has already
been reviewed~\cite{lyn95,tmg96,Perinova:1998ve,Barnett:2007lq}.
 
Although the definition of the absolute phase is in itself an
interesting problem, such an absolute phase has no meaning from a
practical point of view. Strictly speaking, only relative variables
are of interest in physics. Most, if not all, methods of phase
measurement are arrangements determining the relative phase between
two different modes. One might falsely expect that the relative phase
should be constructed merely as the difference of phases. Perhaps
surprisingly, experience demonstrates that this is not the case: there
are theoretical and experimental results that cannot be accounted for
using the difference of phases~\cite{ls96}. In view of this, one
should start a study of the relative phase without any previous
assumption about single-mode phases.  In particular, the conjugate
variable to a relative phase is a number difference that is not
bounded from below. Thus, it is reasonable to expect that the relative
phase will be free of the problems arising in the one-mode case.

In the characterization of the relative phase for two-mode fields, the
Stokes parameters play an important role~\cite{Luis:2000vn}. From an
experimental point of view, they are measurable quantities. An
essential observation is that they are formally also elements of
$\mathfrak{su}(2)$, which turns out to be the dynamical symmetry
algebra of a qubit, in the modern parlance of quantum
information~\cite{Chuang:2000}.  Interesting links with finite quantum
systems have been discussed in
references~\cite{Daoud:2010fr,Kibler:2008mz}.  In fact, as shown time
ago in the pioneering works of L\'evy-Leblond~\cite{lev73} and
Vourdas~\cite{Vou90,Vourdas:1993bh,Vourdas:1993fj} (see also
\cite{ce90}), the polar decomposition of $\mathfrak{su}(2)$ operators
gives rise to a \textit{bona fide} phase operator, which is also
complementary to the population difference, generalizing somehow Dirac
original idea~\cite{ls93c,sl94,Luis:1997fj}.

One might believe ---again falsely--- that the passage from two-level
systems and $\mathfrak{su}(2)$ to three-level systems and
$\mathfrak{su}(3)$ would be immediate.  We realize in this paper that
this is not so: there appears to be no general phase operator for
$\mathfrak{su}(3)$ that simultaneously verify a polar decomposition,
hermiticity and adequate commutation relations.

The polar decomposition of an operator is always possible in any
dimension (although it is generally not unique). On the other hand,
complementarity in finite dimensional spaces is usually implemented
via finite Fourier transformations.  It will be shown that the happy
coincidence where both concepts occur in the same problem must be in
general abandoned for $\mathfrak{su}(3)$ and more generally for
$\mathfrak{su}(n)$: barring exceptional circumstances, complementarity
and polar decompositions are apparently incompatible.

Because it is generalized much more easily, the definition of phase
operators obtained via polar decompositions will be studied in details,
with special attention to the Lie algebra $\mathfrak{su}(3)$. For the
general $\mathfrak{su}(n)$ case, it is natural to define $n-1$
relative phase operators; we find that they do not in general commute,
except in very specific circumstances where the dimension of the
system is $n$ or goes to infinity.  It will be shown how to quantify
this lack of commutativity and how this can be related to simple
counting arguments based on the geometry of $\mathfrak{su}(n)$ weight
space.

We have found that considerable insight in the structure of phase
operators, emphasizing the connection between polar decomposition
methods, the abstract operators and the geometrical nature of the
relative phase, is gained by introducing a coherent-state realization
of the generators (see
references~\cite{Guise:2002rt,GarciadeLeon:2007ly} for variations on
this theme).  These realizations also provide very useful
calculational simplifications, particularly as representations become
large and as we increase the rank $n-1$ of $\mathfrak{su}(n)$.  We
introduce this representation first for $\mathfrak{su}(2)$, in section
\ref{su2phaseop}, reserving a wealth of mathematical details for
\ref{repGamma}.  In particular, these realizations allow us to reach
some conclusions about the commutativity of phase operators for
$\mathfrak{su}(n)$ with $n\ge 3$, in the limit of large
representations.  This limit can be seen as a classical limit, thus
allowing our conclusions to be checked against classical concepts
associated with relative phases.

\section{$SU(2)$ phase operators}\label{su2phaseop}

The complex extension of the $\mathfrak{su}(2)$ algebra is generated
by the operators $\{ \op{h}, \op{e}_{+} ,\op{e}_{-} \}$ with
commutation relations
\begin{equation}
  \label{eq:ccrsu2}
  [\op{h}, \op{e}_{\pm}] = \pm  \op{e}_{\pm} \, ,
  \qquad 
  [\op{e}_{+}, \op{e}_{-}] = 2 \op{h} \, . 
\end{equation}
In the $(2j+1)$-dimensional space $V_{j}$ spanned by the vectors $\{ |
j m \rangle : \, m = -j, \ldots, j\}$, which is the carrier of the
irreducible representation (irrep) with spin $j$, the generators act
in the standard way:
\begin{equation}
  \label{eq:su2sub}
  \op{h} |jm \rangle = m |jm \rangle \, , 
  \qquad 
  \op{e}_{\pm}  | j m \rangle = \sqrt{(j \mp m)(j \pm m +1)} |j m\pm 1
  \rangle \, .
\end{equation}
The state $| \chi_{j} \rangle \equiv |j j \rangle$ is the highest weight of
the irrep, so that
\begin{equation}
  \label{eq:ms}
  \op{h} | \chi_{j} \rangle = j  | \chi_{j} \rangle \, , 
  \qquad 
  \op{e}_{+} | \chi_{j} \rangle = 0   \, . 
\end{equation}

An advantageous realization providing a link with the two-mode
relative phase, is given by the Schwinger realization of
$\mathfrak{su}(2)$ in terms of two bosonic fields $\op{a}_{1}$ and
$\op{a}_{2}$~\cite{Schwinger:1965kx,Chaturvedi:2006vn}:
\begin{equation}
  \label{eq:Srep}
  \op{e}_{+} \mapsto \op{a}_1^\dagger \, \op{a}_2\, ,
  \qquad
  \op{e}_{-} \mapsto \op{a}_2^\dagger \, \op{a}_1 \, ,
  \qquad
  \op{h} \mapsto \textstyle\frac{1}{2}  ( \op{a}_1^\dagger \op{a}_1 - 
  \op{a}_2^\dagger \op{a}_2  ) \, ,
\end{equation}
which are the building blocks of the Stokes
operators~\cite{Luis:2000vn}.  They act on the two-dimensional
harmonic oscillator basis $|n_1,n_2\rangle $ related to the angular
momentum states $|j m \rangle$ by $n_1+n_2=2j$ and $n_1-n_2=2m$.

If, for a moment we interpret $\op{a}_{1}$ and $\op{a}_{2}$ as
classical field amplitudes, it is apparent from (\ref{eq:Srep}) that
the relative phase between the fields is encoded in
$\op{e}_{\pm}$. Therefore, it seems natural enough to look for a polar
decomposition of the ladder operators~\cite{Lancaster:1985vn}
\begin{equation}
  \label{eq:poldec}
  \op{e}_{-} = \op{E}\, \op{D} \, , 
\end{equation}
where, using (\ref{eq:su2sub}), $\op{D} = \sqrt{\op{e}_{-}^{\dagger}
  \op{e}_{-}}$ is a semi-positive self-adjoint operator and $\op{E}$
is a unitary operator that can be interpreted as the exponential of a
putative phase operator.
  
The rank of $\op{e}_{-}$ is one less than the dimension of
$\op{e}_{-}$, so $\op{E}$ is not completely specified.  We remove the
ambiguity by using cyclic boundary conditions so that $\op{E}$ is the
generator of an Abelian cyclic group~\cite{Luis:1997fj}.  The
eigenvalues of $\op{E}$ are then the quantized phases. As $\op{E}^{ 2
  j +1} = \op{\openone}$, these eigenvalues are just $\omega^{k}$,
where $\omega =\exp[ 2 \pi i/(2 j +1)] $ and $ k=-j, \ldots, j$. 
Taking $m$ modulo $2j+1$, we
get in this way
\begin{equation}
  \op{E} = \sum_{m=-j}^j | j {m-1}\rangle \langle j m | \, , 
\end{equation}
or, equivalently,
\begin{equation}
  \op{E} = 
  \left(
    \begin{array}{cccccc}
      0 & 0 & \ldots &  & 0 & 1 \\
      1 & 0 &  &  &  &  \\
      & 1 & 0 &  &  &  \\
      &  &  & \ddots &  &  \\
      &  &  & 1 & 0 &  \\
      &  &  &  & 1 & 0 
    \end{array}
  \right) \, .
\end{equation}
The eigenstates of $\op E$ are related to the basis states $\vert
jm\rangle$ by a finite Fourier transform and are thus complementary to
the $\vert jm\rangle$ states.

Observe that $\op{E}$ is not an $SU(2)$ matrix.  It can be written
formally as the exponential of a Hermitian phase operator
$\op{\varphi}$, but $\op{\varphi}$ is not in general an element of the
$\mathfrak{su}(2)$ algebra.  These observations hold (even in
dimension $2$ because of the choice of phase in completing $\op{E}$),
even though both have well-defined actions on the basis elements.

As heralded in the introduction, we now introduce as a tool of
particular convenience a coherent state realization $\Gamma$ for
$\mathfrak{su}(2)$.  $\Gamma$
is not Hermitian, although it is equivalent by similarity
transformation to the ``standard'' Hermitian realization of
(\ref{eq:su2sub}), as shown in \ref{repGamma}.

The $SU(2)$ coherent states are defined
by~\cite{Arecchi:1972fk,Perelomov:1986,Gazeau:2009kx}
\begin{equation}
  \label{eq:defCS}
  | \vartheta, \varphi \rangle = 
  \op{R}_{z}^{-1} (\varphi) \op{R}_{y}^{-1}  (\vartheta)  | \chi_{j} \rangle \, ,
\end{equation}
where $\op{R}_{z}$ and $\op{R}_{y}$ represent rotation about the $z$
and $y$ axes, respectively. To any vector $| \Psi \rangle$ we
associate the function
\begin{equation}
  | \Psi \rangle  \mapsto  \Psi_{\vartheta} (\varphi) = 
  \langle \chi_{j} |  \op{R}_{y}  (\vartheta)  \, \op{R}_{z} (\varphi) |
  \Psi \rangle\, .
\end{equation}
Note that $| \Psi_{\vartheta} (\varphi) |^{2}$ is precisely the Husimi
$Q$-function for the corresponding (pure) state $| \Psi\rangle $.  We
can use the arbitrary nature of $| \Psi \rangle$ to define the action
of $\op{X}\in \mathfrak{su}(2)$ on $\Psi_\vartheta(\varphi)$ by:
\begin{equation}
  \op{X} | \Psi \rangle  \mapsto  
  [  \Gamma ( \op{X} )  \Psi ]_{\vartheta}(\varphi) \equiv  
  \langle \chi_{j} |  \op{R}_{y}  (\vartheta) 
  \op{R}_{z} (\varphi) \op{X} | \Psi \rangle \, . \label{Gammamap}
\end{equation}
Straightforward manipulations immediately produce the expressions
\begin{eqnarray}
  \label{torus2}
  \op{h}   \mapsto \Gamma ( \op{h} ) = -i \frac{d}{d\varphi} \, ,
  \nonumber \\
  &  & \\
  \op{e}_{\pm} \mapsto \Gamma ( \op{e}_{\pm} ) = 
  -( \tan\vartheta)^{\mp 1} \, \e^{\pm i\varphi } 
  \left (  j \mp  i \frac{d}{d\varphi} \right ) \, . \nonumber
\end{eqnarray}
It is convenient to think of the coherent state $ | \vartheta, \varphi
\rangle $ as localized around the coordinates $(\vartheta,\varphi)$ on
the Bloch sphere. The relative phase is linked to the azimuthal angle
$\varphi$ while the parameter $\vartheta$ is inessential for our
purposes; to simplify the expressions for $\Gamma$, we choose
$\vartheta$ so that $\tan \vartheta = -1$ and obtain
\begin{eqnarray}
  \label{torus2}
  \op{h}   \mapsto  \Gamma ( \op{h} ) = -i \frac{d}{d\varphi} \, ,
  \nonumber \\
  &  & \\
  \op{e}_{\pm} \mapsto \Gamma ( \op{e}_{\pm} ) =
  \e^{\pm i\varphi } \left (  j \mp  i \frac{d}{d\varphi} \right ) \, . \nonumber
\end{eqnarray}

One easily verifies that $\hat X\mapsto \Gamma(X)$ preserves the
commutation relations (\ref{eq:ccrsu2}) and is thus a realization of
$\mathfrak{su}(2)$.  $\Gamma$ acts naturally in the
infinite--dimensional space spanned by the exponential functions $ \{
\e^{im \varphi} : \, 2m \in \mathbb{Z} \}$ and equipped with scalar
product
\begin{equation}
  \label{inner}
  ( f | g )=\int _0^{2\pi}  f^{\ast} (\varphi) \,
  g(\varphi) \, d\varphi \, . \label{iponthetorus}
\end{equation}
The subspace of states with $| m | \le j$ is invariant under the
action of $\Gamma$.  The (normalized) basis elements 
of this invariant subspace are mapped to
exponential functions $| jm \rangle \leftrightarrow \exp
(im\varphi)/\sqrt{2 \pi}$ and the action of (\ref{torus2}) is
\begin{equation}
  \Gamma( \op{h} ) | jm \rangle    = m | j m \rangle
  \qquad 
  \Gamma (\op{e}_{\pm} )  | jm \rangle  = ( j \mp m )\, |j m \pm 1 \rangle  \, .   
  \label{100}
\end{equation}
Under this inner product
\begin{equation}
  \langle j m^\prime | \Gamma( \op{e}_{+} ) |  jm\rangle  \neq
  \langle jm^\prime | \Gamma^\dagger ( \op{e}_{-})  | jm\rangle \, ,
  \label{101}
\end{equation}
thus, the realization $\Gamma$ is not Hermitian.  However, we show in
\ref{repGamma} how, for fixed $j$, $\Gamma$ is equivalent to the
standard Hermitian representation given in equation~(\ref{eq:su2sub}).

The considerable merit of the realization $\Gamma$ is that it is
particularly well-suited to analyze the polar decomposition: the
unitary matrix $\op{E}$ in $V_j$ is immediately obtained from the
action of $\e^{-i\varphi}$, the ``phase'' part of $\Gamma(\hat e_-)$:
\begin{equation}
  \op E_{m'm}=\Gamma( \op{E} )_{m'm}= \langle j m' |  \e^{-i\varphi} |
  jm\rangle \, ,  \qquad
  -j \le m\le j \, \quad \hbox{\rm mod}(2j+1)\, .
\end{equation}
Thus, $\e^{-i\varphi}$ simply shifts the basis state $\e^{im\varphi}$
on the circle to its immediate neighbour $\e^{i(m-1)\varphi}$, modulo
$2j+1$.  In addition, matrix elements of $\Gamma$ becomes
indistinguishable from those of the standard hermitian representation
in the limit where $m/j\to 0$: this {makes $\Gamma$ also very well
  suited to analyze some limits of large representations, and analyze
  a transition between the quantum and classical phase.

  In the basis of exponential functions, the $k$'th eigenstate
  $|\Phi_{k}\rangle $, corresponding to the eigenvalue $\omega^{k}$,
  is (up to an overall phase)
  \begin{equation}
    \langle \varphi \,|\,\Phi_{k}\rangle =
    \frac{1}{\sqrt{2j+1}} \ 
    \frac{\sin [( 2 j+1) \varphi ] }
    {\sin \left ( \varphi +  \frac{\pi  k}{2j +1} \right ) } \, .
  \end{equation}

  \section{$SU(3)$ phase operators}

  \subsection{Polar decomposition for $SU(3)$}

  A basis for the complex extension of the Lie algebra
  $\mathfrak{u}(3)$ is given by the nine operators $\{\op{C}_{ij} : \,
  i,j=1,2,3\} $, with commutation relations
  \begin{equation}
    [ \op{C}_{ij}, \op{C}_{k\ell} ]= \delta_{jk} \, \op{C}_{i\ell} - 
    \delta_{i\ell} \, \op{C}_{kj} \, .  
    \label{u3commute}
  \end{equation}
  The complex extension of $\mathfrak{su}(3)$ is obtained by
  restricting the $\mathfrak{u}(3)$ operators to $\{ \op{C}_{ij}: \; i
  \neq j \}$ and including two traceless linearly independent diagonal
  operators $\op{h}_{1}$ and $\op{h}_{2}$ that determine a Cartan
  subalgebra.  A convenient choice of the latter is
  \begin{equation}
    \op{h}_{1}  = \op{C}_{11} - \op{C}_{22} \, , 
    \qquad
    \op{h}_{2} =\op{C}_{22} - \op{C}_{33} \, .
  \end{equation}
  If one uses the boson realization:
  \begin{equation}
    \op{C}_{ij} = \op{a}_{i}^{\dag}  \op{a}_{j} \, , 
    \qquad 
    \op{C}_{ij} = \op{C}_{ji}^\dagger \, , 
    \label{su3boson}
  \end{equation}
  the $\textstyle{\frac{1}{2}} (\lambda+1)(\lambda+2)$-dimensional set
  of harmonic oscillator states $\mathcal{S} = \{ | n_1 n_2 n_3\rangle
  , n_1+n_2+n_3=\lambda\}$, is left invariant under the action of
  $\op{C}_{ij}$ and is a basis for an irrep of $\mathfrak{su}(3)$
  usually denoted $(\lambda, 0)$.  The eigenvalues of $\op{h}_{1}$ and
  $\op{h}_{2}$ are directly related to population differences between
  levels 1 and 2, and 2 and 3, respectively.

  The information in equation~(\ref{u3commute}) is conveniently
  displayed using a root diagram~\cite{Hall:2003ul}: to every
  $\mathfrak{su}(3)$ generator $\op{C}_{ij}$ we associate the pair
  $(x, y)$ of integers defined by
  \begin{equation} [ \op{h}_{1} , \op{C}_{ij} ] = x \ \op{C}_{ij} \, ,
    \qquad [ \op{h}_{2} , \op{C}_{ij} ] = y \ \op{C}_{ij} \, ,
  \end{equation}
  and the root vector $\alpha= x \, \alpha_{1} + y \, \alpha_{2}$,
  with basis $\alpha_1, \alpha_2$ having Cartesian components
  \begin{equation}
    \alpha_{1} = \left ( \sqrt{2}, 0 \right ) \, ,
    \qquad 
    \alpha_{2} = \left (- \frac{\sqrt{2}}{2}, \frac{\sqrt{6}}{2} \right ) \, .
  \end{equation}
  The root diagram for $\mathfrak{su}(3)$ is sketched in
  figure~\ref{su3root}.  Note that every generator is thus associated
  with a root vector and the diagonal operators are associated with
  vectors of length zero.

  \begin{figure}
    \centerline{\includegraphics[scale=0.35]{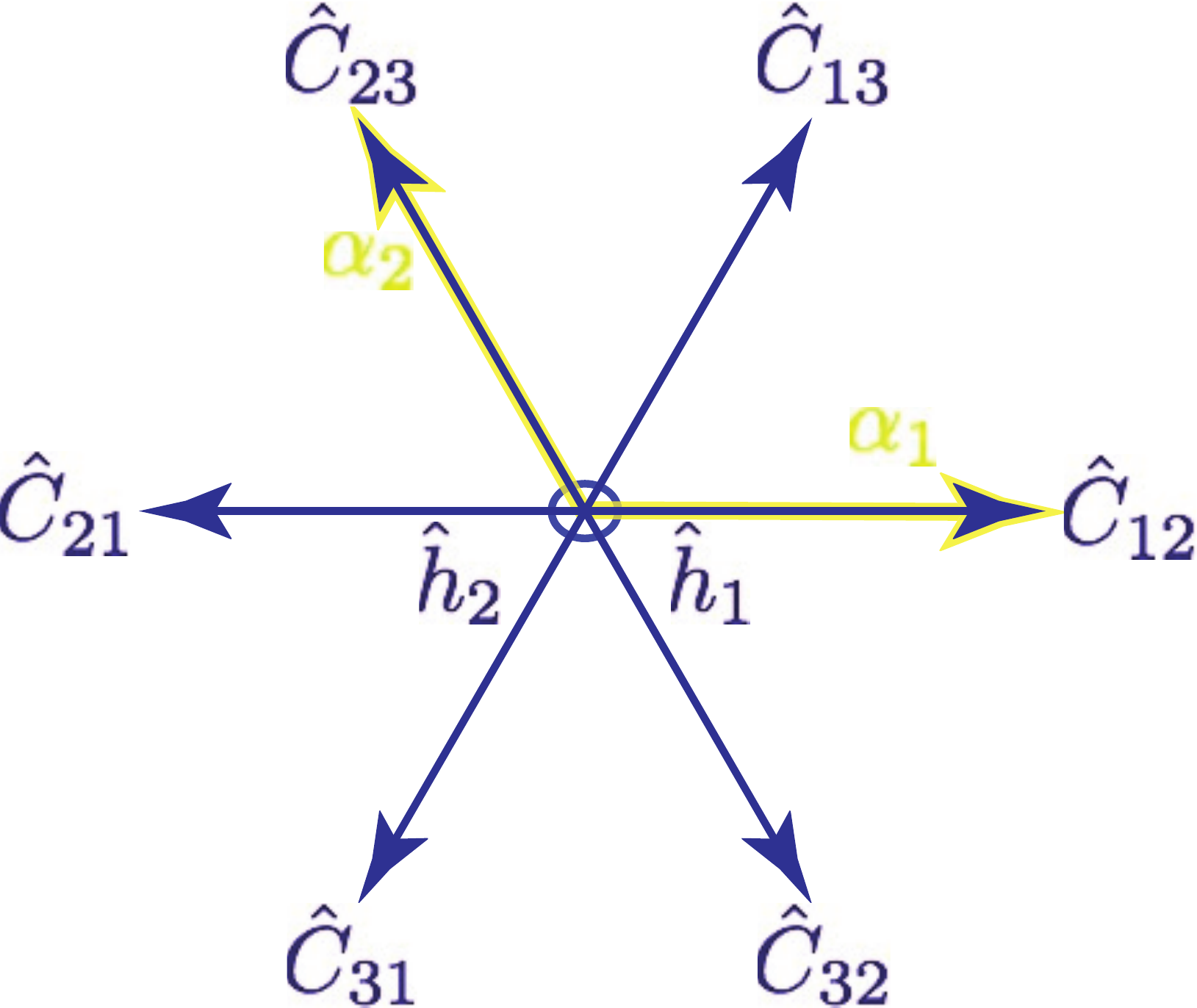}}
    \caption{The root system for the complex extension of
      $\mathfrak{su}(3)$, showing (in yellow) the two fundamental
      positive roots.}
    \label{su3root}
  \end{figure}

  The commutation relations are given (up to a sign) by addition of
  the corresponding root vectors. If we label the operator
  $\op{C}_{ij}$ by its root vector, $\op{C}_{ij}\mapsto
  \op{e}_{\alpha}$, then we have
  \begin{equation}
    [ \op{e}_{\alpha} , \op{e}_{\beta} ] \propto 
    \left\{
      \begin{array}{ll}
        \op{e}_{\alpha +\beta }, & 
        \hbox{if  $\alpha +\beta$  is a  root,} \\
        & \\
        0, &  \hbox{ otherwise.}
      \end{array}
    \right .
  \end{equation}
  The root diagram neatly shows, for instance, that $ [\op{C}_{23},
  \op{C}_{12}]$ is proportional to $\op{C}_{13}$ in accordance to the
  vectorial addition of the appropriate roots.

  Similarly, the weight diagram is a pictorial representation of the
  basis states of an irrep. The weight $w=x \, w^1+y \, w^2$ of a
  basis state $| w \rangle $ is a vector with components related to
  the eigenvalues of the diagonal operators:
  \begin{equation}
    \op{h}_{1} | w \rangle = x \ | w \rangle \, ,
    \qquad
    \op{h}_{2} | w \rangle = y \ | w \rangle \, .
  \end{equation} 
  The fundamental weights $w^{1}$ and $w^{2}$ have Cartesian
  coordinates
  \begin{equation}
    w^{1} = \left ( \frac{1}{\sqrt{2}}, \frac{1}{\sqrt{6}} \right ) \, , 
    \qquad
    w^{2} = \left ( 0, \sqrt{\frac{2}{3}} \right ) \, , 
  \end{equation} 
  so that $\langle w^i | \alpha_j \rangle =\delta_{ij}$.  A generator
  $\op{C}_{ij}$ associated with the root $\alpha$ acts on a weight
  state $| w \rangle $ by translation on the hexagonal grid:
  \begin{equation}
    \op{C}_{ij} |  w \rangle \mapsto \op{e}_{\alpha} | w \rangle 
    \propto \left\{
      \begin{array}{ll}
        | \alpha + w \rangle , &  
        \hbox{if $\alpha +w$  is a weight} ,      \\
        & \\
        0, & \hbox{otherwise.} 
      \end{array}
    \right.
  \end{equation}

  The operators $\op{C}_{ij}$ acting on the $\textstyle{\frac{1}{2}}
  (\lambda+1)(\lambda+2)$-dimensional space $\mathcal{S}$ can be
  represented by matrices.  A polar decomposition of these matrices is
  given by
  \begin{eqnarray}
    \op{C}_{ij} =  \op{E}_{ij}  \op{D}_{ij} \, .
  \end{eqnarray}
  The operator $\op{D}_{ij} = \sqrt{\op{C}^\dagger_{ij}\op{C}_{ij}} $
  is non-negative definite while $\op{E}_{ij }$ will be constructed as
  unitary matrices, with $\op{E}_{11} = \op{E}_{22} =
  \op{E}_{33}=\openone$.  It is easily shown that $[ \op{h}_{1},
  \op{D}_{ij}] = [\op{h}_{2} , \op{D}_{ij}]=0$ and also that
  \begin{equation}
    \op{D}_{12}^2  - \op{D}_{21}^2 = \op{h}_{1}  \, , 
    \qquad
    \op{D}_{23}^2 - \op{D}_{32}^2 = \op{h}_{2}  \, , 
    \qquad
    \op{D}_{13}^2 - \op{D}_{31}^2 = \op{h}_{1} + \op{h}_{2} \, .
  \end{equation}

  The realization of $\mathfrak{su}(3)$ that optimally displays the
  polar decomposition is the $\mathfrak{su}(3)$ analogue of
  (\ref{torus2}).  A coherent state
  $|\vartheta_1,\varphi_1,\vartheta_2,\varphi_2\rangle$ for the irrep
  $(\lambda,0)$ is obtained by group action on the highest weight
  state; this highest weight state is the boson state $| \lambda, 0,
  0\rangle$.  A general quantum state $| \Psi\rangle$ is then
  represented by a function on $S^4\sim SU(3)/U(2)$
  \begin{equation}
    | \Psi\rangle\mapsto \Psi_{\vartheta_1, \vartheta_2}(\varphi_1,\varphi_2)
  \end{equation}
  The two azimuthal angles $\varphi_1,\varphi_2$ control the relative
  phase between these populations, while two polar angles
  $\vartheta_1,\vartheta_2$ mix the number of excitations in each
  mode.  For calculational convenience the latter are chosen to
  simplify the coherent state representation of $\frak{su}(3)$
  elements, given by~\cite{Guise:2002rt}
  \begin{eqnarray}
    \label{Gammasu3}
    \op{h}_{1} & \mapsto & \Gamma ( \op{h}_{1} ) =  
    -i \frac{\partial}{\partial \varphi_{1}} \, , 
    \qquad
    \op{h}_{2}  \mapsto  \Gamma ( \op{h}_{2} ) =  
    -i \frac{\partial}{\partial \varphi_{2}} \, , \nonumber \\
    \op{C}_{12} & \mapsto  & \Gamma ( \op{C}_{12} ) = 
    \frac{1}{3} \e^{i(2\varphi_{1}-\varphi_{2})}
    \left (  \lambda +i \frac{\partial }{\partial \varphi_{1}} 
      - i \frac{\partial}{\partial \varphi_{2}}  \right ) \, ,\\
    \op{C}_{23} & \mapsto & \Gamma ( \op{C}_{13} ) = 
    \frac{1}{3} \e^{-i(\varphi_{1} -2 \varphi_{2})}
    \left (  \lambda + i\frac{\partial }{\partial \varphi_{1}} 
      +2 i \frac{\partial }{\partial \varphi_{2}}
    \right )  \, . \nonumber
    \label{su3polar}
  \end{eqnarray}
  This realization provides an obvious decomposition of the
  $\mathfrak{su}(3)$ raising operators.  Much like $\mathfrak{su}(2)$,
  these operators act in a natural way on the Hilbert space spanned by
  the exponential functions $\{
  e^{i(w_{1}\varphi_{1}+w_{2}\varphi_{2})} \}$. Again, $\Gamma$ is not
  Hermitian although it is equivalent to a Hermitian representation,
  as indicated in Reference~\cite{Guise:2002rt} or as can 
  be shown following the procedure of \ref{repGamma}.

  \subsection{$SU(3)$ phase operators for the (1,0) representation}

  We consider first the three-dimensional representation $(1,0)$,
  spanned by the boson states $\{ | 100 \rangle ,| 010 \rangle, | 001
  \rangle \}$, where $|n_{1} n_{2} n_{3} \rangle $ denotes a state
  with population $n_{i}$ in level $i$. This has been worked out in
  detail in~\cite{Klimov:2004ly} and from a different perspective
  in~\cite{Daoud:2011ul}.

  The components $(x,y)$ of a weight are related to population
  differences by $ x = n_1 -n_2, y=n_2-n_3$.  Explicitly, the weights
  of the basis vectors are $\{(1,0), (-1,0),(0,-1)\}$, respectively.

  In this representation, we have the matrix realizations
  \begin{eqnarray}
    \op{ C}_{12} & = & 
    \left (
      \begin{array}{ccc}
        0 & 1 & 0 \\
        0 & 0 & 0 \\
        0 & 0 & 0
      \end{array}
    \right) = 
    \op{E}_{12}
    \left(
      \begin{array}{ccc}
        0 & 0 & 0 \\
        0 & 1 & 0 \\
        0 & 0 & 0
      \end{array}
    \right)\, , \nonumber \\
    & & \\
    \op{C}_{23} & = & 
    \left(
      \begin{array}{ccc}
        0 & 0 & 0 \\
        0 & 0 & 1 \\
        0 & 0 & 0
      \end{array}
    \right) = 
    \op{E}_{23} 
    \left(
      \begin{array}{ccc}
        0 & 0 & 0 \\
        0 & 0 & 0 \\
        0 & 0 & 1
      \end{array}
    \right)\, . \nonumber 
  \end{eqnarray}
  The rank of $\op{C}_{12}$ and $\op{C}_{23}$ is one, but their
  dimension is three, which implies that the polar decomposition is
  (again) not completely specified.  Indeed, one finds that the most
  general unitary $\op{E}_{12}$ and $\op{E}_{23}$ consistent with the
  matrix realization of $\op{C}_{12}$ and $\op{C}_{23}$ are
  \begin{eqnarray}
    \label{e23}
    \op{E}_{12}& = & \left(
      \begin{array}{ccc}
        0 & 1 & 0 \\
        a & 0 & b \\
        b^{\ast } & 0 & -a^{\ast}
      \end{array}
    \right ) \, , 
    \qquad  a a^{\ast} + b b^{\ast} = 1 \, ,   
    \nonumber \\
    & & \\
    \op{E}_{23} & = & \left(
      \begin{array}{ccc}
        c & d & 0 \\
        0 & 0 & 1 \\
        d^{\ast } & -d^{\ast } & 0
      \end{array}
    \right) \, , 
    \qquad c c^{\ast }+ d d^{\ast} = 1 \, . 
    \nonumber  
  \end{eqnarray}
  Here, we have already restricted $\op{E}_{ij}$ to be unitary so that
  a {Hermitian phase operator can be properly defined.  The issue is
    now to fix the unknown parameters $a, b, c$ and $d$ in
    (\ref{e23}).

    \begin{table}
      \caption{\label{blobs} Notational details connecting the numbering of states,
        their weights, their boson representations, and the expression of these states
        in terms of the angles $\varphi_1,\varphi_2$. In the polar
        representation we omit for simplicity a factor $2\pi$.}
      \begin{indented}
      \item[]\begin{tabular}{@{}lcll} \br
          State & Weight & Boson state & Polar state \\
          \mr $| 1 \rangle $ & $ ( 1, 0 )$ & $ | 100 \rangle$ &
          $ \e^{i\varphi_{1}}$ \\
          $| 2 \rangle $ & $ ( -1 , 1 )$ & $ | 010 \rangle$ &
          $ \e^{i ( -\varphi_{1} + \varphi_{2})}$ \\
          $| 3 \rangle $ & $ ( 0 , -1 )$ & $ | 001 \rangle$ &
          $ \e^{- i \varphi_{2}}$ \\
          \br
        \end{tabular}
      \end{indented}
    \end{table}

    \subsubsection{$SU(2)$-invariant solution.}

    The subset $\{ \op{C}_{12}, \op{C}_{21}, \op{h}_{1} \} $ of
    generators spans an $\mathfrak{su}(2)$ subalgebra of
    $\mathfrak{su}(3)$. The boson states $| 100\rangle $ and $|
    010\rangle $ form a two-dimensional $\mathfrak{su}(2)$ subspace;
    the boson state $| 001\rangle $ is an $\mathfrak{su}(2)$ singlet.
    Thus, one way of fixing the unitary matrix $\op{E}_{12}$ is to
    require that $\op{E}_{12}$ preserve the multiplet structure of
    this $\mathfrak{su}(2)$ subalgebra. This gives $a= -1, b=0$, so
    that
    \begin{equation}
      \op{E}_{12} =
      \left (
        \begin{array}{rrr}
          0 & 1 & 0 \\
          -1 & 0 & 0 \\
          0 & 0 & 1%
        \end{array}
      \right) \, .  
      \label{su212}
    \end{equation}
    The phase of $a$ is inessential and has been chosen for
    convenience.

    In the same way, the subset $\{ \op{C}_{23}, \op{C}_{32},
    \op{h}_{2} \}$ spans a different $\frak{su}(2)$ subalgebra, and we
    may also require that $\op{E}_{23}$ act within a multiplet of this
    subalgebra. This, in turn, implies
    \begin{equation}
      \op{E}_{23} = \left(
        \begin{array}{rrr}
          1 & 0 & 0 \\
          0 & 0 & 1 \\
          0 & -1 & 0
        \end{array}
      \right) \,.  \label{su223}
    \end{equation}
    For this solution, one obtains, from
    $\op{E}_{ij}=\e^{i\op{\varphi}_{ij}}$,
    \begin{equation}
      \op{\varphi}_{12} = i {\frac{\pi }{2}}
      \left(
        \begin{array}{rrr}
          0 & -1 & 0 \\
          1 & 0 & 0 \\
          0 & 0 & 0
        \end{array}
      \right) \, ,
      \qquad 
      \op{\varphi}_{23} = i{\frac{\pi }{2}}
      \left(
        \begin{array}{ccc}
          0 & 0 & 0 \\
          0 & 0 & -1 \\
          0 & 1 & 0
        \end{array}
      \right) \, . 
      \label{phiij}
    \end{equation}
    The notable feature of the matrices $\op{E}_{12}$ and
    $\op{E}_{23}$ is that they do not commute.

    \subsubsection{The complementary solution.}

    An alternative choice of $\op{E}_{ij}$ is obtained by using a
    different line of argument. For $SU(2)$, the phase operator is
    thought to be complementary to the population difference
    $\op{h}$. The generalization of this complementarity-based
    definition to $SU(3)$ can also be achieved for the irrep $(1,0)$.

    We recall the definition of the generalized Pauli
    matrices~\cite{Schwinger:1960a,Schwinger:1960b,Schwinger:1960c}. Let
    \begin{equation}
      \op{Z} =\op{h}_{1} = 
      \left(
        \begin{array}{ccc}
          \omega & 0 & 0 \\
          0 & \omega ^{2} & 0 \\
          0 & 0 & 1
        \end{array}
      \right ) \, ,
      \qquad  
      \op{X} = \left (
        \begin{array}{ccc}
          0 & 1 & 0 \\
          0 & 0 & \omega^{2} \\
          \omega & 0 & 0
        \end{array}
      \right) \, .
    \end{equation}
    Then
    \begin{equation}
      \op{X}^k \op{Z}^\ell = \omega^{k\ell} \op{Z}^\ell \op{X}^k \, ,
    \end{equation}
    where $k,\ell\in \mathbb{Z}_{3}$ and $\omega=\exp (2 \pi i/3)$.
    The subset $\{\op{X}^k \op{Z}^\ell\}$ of generalized Pauli
    matrices are elements of the finite Pauli subgroup $\wp_{3}$ of
    $SU(3)$ containing 27 elements and described
    elsewhere~\cite{Patera1988,Patera1989}.  $\{\op{X}^k
    \op{Z}^\ell\}$ also forms a basis for the $\mathfrak{su}(3)$
    algebra, so that we can expand $\op{E}_{12}$ as
    \begin{equation}
      \op{E}_{12} = \sum_{k\ell} a_{k\ell} \op{X}^k \op{Z}^\ell\, .
    \end{equation}
    A simple analysis shows that it is indeed possible to obtain a
    complementary solution, in the sense that
    \begin{equation}
      \op{h}_{1} \op{E}_{12}  = \omega^{2} \op{E}_{12} \op{h}_{1} \, ,
    \end{equation}
    which then forces
    \begin{equation}
      \op{E}_{12} =\left(
        \begin{array}{ccc}
          0 & 1 & 0 \\
          0 & 0 &\e^{i\beta } \\
          \e^{-i\beta } & 0 & 0
        \end{array}
      \right) \, .
    \end{equation}
    A similar analysis for $\op{E}_{23}$ and $\op{h}_{2}$ produces the
    unitary solution
    \begin{equation}
      \op{E}_{23} =
      \left(
        \begin{array}{ccc}
          0 & \e^{i\gamma } & 0 \\
          0 & 0 & 1 \\
          \e^{-i \gamma } & 0 & 0
        \end{array}
      \right) \, .
    \end{equation}

    If, additionally. we insist that the $\op{E}_{ij}$'s commute or,
    equivalently, $\op{\varphi}_{13} = \op{\varphi}_{12} +
    \op{\varphi}_{23}$, we obtain the conditions $\op{E}_{13} =
    \op{E}_{12} \, \op{E}_{23}$, which implies
    \begin{equation}
      \beta +\gamma = 0, \pm 2\pi , \ldots 
      \quad
      2 \beta -\gamma  = 0,\pm 2\pi , \ldots 
      \quad 
      -\beta + 2\gamma =0,\pm 2\pi , \ldots \, ,
    \end{equation}
    wherefrom we get
    \begin{equation}
      3 \beta =0,\pm 2\pi ,\ldots\, , 
      \qquad 
      3 \gamma =0,\pm 2\pi ,\ldots \, .
    \end{equation}
    The solutions are found by fixing either $\beta$ and deducing
    $\gamma$, or visa versa. The simplest nontrivial solution is found
    by choosing $\beta =2\pi /3$ and $\gamma =-2\pi /3$. This produces
    \begin{equation}
      \fl
      \op{E}_{12}= \left (
        \begin{array}{ccc}
          0 & 1 & 0 \\
          0 & 0 & \omega \\
          \omega ^{2} & 0 & 0%
        \end{array}
      \right), 
      \quad  
      \op{E}_{23} =\left (
        \begin{array}{ccc}
          0 & \omega ^{2} & 0 \\
          0 & 0 & 1 \\
          \omega & 0 & 0
        \end{array}
      \right ), 
      \quad 
      \op{E}_{13} = \left (
        \begin{array}{ccc}
          0 & 0 & 1 \\
          \omega ^{2} & 0 & 0 \\
          0 & \omega & 0
        \end{array}
      \right ) \, .
    \end{equation}
    all of which are elements of the generalized Pauli group
    $\wp_{3}$.

    \subsection{$SU(3)$ phase operators in the $( \lambda, 0) $
      representation}

    We now seek to generalize our discussion to irreps other than the
    simplest $(1,0)$.  The $d=\frac{1}{2} ( \lambda +1 ) ( \lambda +2
    ) $ states of the form $|n_{1} n_{2} n_{3}\rangle$, where each
    $n_{i}$ is a non-negative integer subject to the condition $n_{1}
    + n_{2} + n_{3}=\lambda $, are transformed into one another under
    the action of any $\mathfrak{su}(3)$ generator and thus form an
    irrep of dimension $d$.  The highest weight is $ ( \lambda , 0 ) $
    and the highest weight state is $| \lambda 0 0\rangle $.  We can
    again use the polar realization of equation~(\ref{su3polar}) and
    express basis states of $(\lambda ,0)$ in terms of
    exponentials. Some closely related material has been presented
    in~\cite{Zanette:2012fk}.

    Consider for instance the case $\lambda=2$, where the dimension of
    the space is $6$.  The operators $\op{C}_{12}$ and $\op{C}_{23}$
    have matrix representation and polar decomposition of the general
    form
    \begin{eqnarray}
      \op{C}_{12} &=& 
      \left(
        \begingroup
        \everymath{\scriptstyle}
        \small
        {\renewcommand{\arraystretch}{0.75}
          \renewcommand\arraycolsep{0.2em}
          \begin{array}{cccccc}
            0 & \sqrt{2} & 0 & 0 & 0 & 0 \\
            0 & 0 & 0 & 1 & 0 & 0 \\
            0 & 0 & 0 & 0 & 1 & 0 \\
            0 & 0 & 0 & 0 & 0 & 0 \\
            0 & 0 & 0 & 0 & 0 & 0 \\
            0 & 0 & 0 & 0 & 0 & 0
          \end{array}
        }
        \endgroup
      \right)
      = \op{E}_{12}
      \left(
        \begingroup
        \everymath{\scriptstyle}
        \small
        {\renewcommand{\arraystretch}{0.75}
          \renewcommand\arraycolsep{0.2em}
          \begin{array}{cccccc}
            0 & 0 & 0 & 0 & 0 & 0 \\
            0 & \sqrt{2} & 0 & 0 & 0 & 0 \\
            0 & 0 & 0 & 0 & 0 & 0 \\
            0 & 0 & 0 & 1 & 0 & 0 \\
            0 & 0 & 0 & 0 & 1 & 0 \\
            0 & 0 & 0 & 0 & 0 & 0
          \end{array}
        }
        \endgroup
      \right),  \nonumber \\ 
      & & \\ 
      \op{C}_{23} &=& 
      \left (
        \begingroup
        \everymath{\scriptstyle}
        \small
        {\renewcommand{\arraystretch}{0.75}
          \renewcommand\arraycolsep{0.2em}
          \begin{array}{cccccc}
            0 & 0 & 0 & 0 & 0 & 0 \\
            0 & 0 & 1 & 0 & 0 & 0 \\
            0 & 0 & 0 & 0 & 0 & 0 \\
            0 & 0 & 0 & 0 & 1 & 0 \\
            0 & 0 & 0 & 0 & 0 & \sqrt{2} \\
            0 & 0 & 0 & 0 & 0 & 0
          \end{array}
        }
        \endgroup
      \right) = 
      \op{E}_{23} 
      \left(
        \begingroup
        \everymath{\scriptstyle}
        \small
        {\renewcommand{\arraystretch}{0.75}
          \renewcommand\arraycolsep{0.2em}
          \begin{array}{cccccc}
            0 & 0 & 0 & 0 & 0 & 0 \\
            0 & 0 & 0 & 0 & 0 & 0 \\
            0 & 0 & 1 & 0 & 0 & 0 \\
            0 & 0 & 0 & 0 & 0 & 0 \\
            0 & 0 & 0 & 0 & 1 & 0 \\
            0 & 0 & 0 & 0 & 0 & \sqrt{2}
          \end{array}
        }
        \endgroup
      \right) \, , \nonumber 
    \end{eqnarray}
    with
    \begin{equation}
      \op{E}_{12} = 
      \left (
        \begingroup
        \everymath{\scriptstyle}
        \small
        {\renewcommand{\arraystretch}{0.75}
          \renewcommand\arraycolsep{0.2em}
          \begin{array}{cccccc}
            0 & 1 & 0 & 0 & 0 & 0 \\
            0 & 0 & 0 & 1 & 0 & 0 \\
            0 & 0 & 0 & 0 & 1 & 0 \\
            \ast & 0 & \ast & 0 & 0 & \ast \\
            \ast & 0 & \ast & 0 & 0 & \ast \\
            \ast & 0 & \ast & 0 & 0 & \ast
          \end{array}
        }
        \endgroup
      \right)\, ,
      \qquad 
      \op{E}_{23} = \left (
        \begingroup
        \everymath{\scriptstyle}
        \small
        {\renewcommand{\arraystretch}{0.75}
          \renewcommand\arraycolsep{0.2em}
          \begin{array}{cccccc}
            \ast & \ast & 0 & \ast & 0 & 0 \\
            0 & 0 & 1 & 0 & 0 & 0 \\
            \ast & \ast & 0 & \ast & 0 & 0 \\
            0 & 0 & 0 & 0 & 1 & 0 \\
            0 & 0 & 0 & 0 & 0 & 1 \\
            \ast & \ast & 0 & \ast & 0 & 0
          \end{array}
        }
        \endgroup
      \right ) \,  .  
      \label{su320}
    \end{equation}
    and the asterisks denoting here undetermined elements.

    \subsubsection{$SU(2)$-invariant solution.}

    One way of fixing $\op{E}_{12}$ and $\op{E}_{23}$ so they are
    unitary is to directly generalize the prescription
    of~(\ref{su212}) and (\ref{su223}) and complete the matrices in an
    $\mathfrak{su}(2)$--invariant way.  The
    $\mathfrak{su}(2)$-invariant solution, which always exists, can be
    obtained by transforming $\mathfrak{su}(2)$-invariant strings of
    weights parallel to a root into a circle, thus transforming the
    equilateral triangle formed by the weights into a cone, as illustrated in figure
    ~\ref{cone}.  The tip of the triangle is an $\mathfrak{su}(2)$
    singlet while the base is made from the longest $\mathfrak{su}(2)$
    string of weights.

    For the case of $(2,0)$}, we obtain
  \begin{equation}
    \op{E}_{12}= \left (
      \begingroup
      \everymath{\scriptstyle}
      {\renewcommand{\arraystretch}{0.75}
        \renewcommand\arraycolsep{0.2em}
        \begin{array}{cccccc}
          0 & 1 & 0 & 0 & 0 & 0 \\
          0 & 0 & 0 & 1 & 0 & 0 \\
          0 & 0 & 0 & 0 & 1 & 0 \\
          -1 & 0 & 0 & 0 & 0 & 0 \\
          0 & 0 & 1 & 0 & 0 & 0 \\
          0 & 0 & 0 & 0 & 0 & 1
        \end{array}
      }
      \endgroup  
    \right ) \,  , 
    \qquad 
    \op{E}_{23} = \left (
      \begingroup
      \everymath{\scriptstyle}
      \small
      {\renewcommand{\arraystretch}{0.75}
        \renewcommand\arraycolsep{0.2em}
        \begin{array}{cccccc}
          1 & 0 & 0 & 0 & 0 & 0 \\
          0 & 0 & 1 & 0 & 0 & 0 \\
          0 & 1 & 0 & 0 & 0 & 0 \\
          0 & 0 & 0 & 0 & 1 & 0 \\
          0 & 0 & 0 & 0 & 0 & 1 \\
          0 & 0 & 0 & -1 & 0 & 0
        \end{array}
      }
      \endgroup  
    \right ) \, .  \label{su320su2}
  \end{equation}
  This $\mathfrak{su}(2)$-preserving solution does not produce
  commuting matrices: the phases are not additive.  This remains true
  for $(\lambda,0)$-type of representations, where the weight diagram
  is an equilateral triangle with $\lambda+1$ states on each side.

  \begin{figure}
    \centerline {\includegraphics[height=5cm]{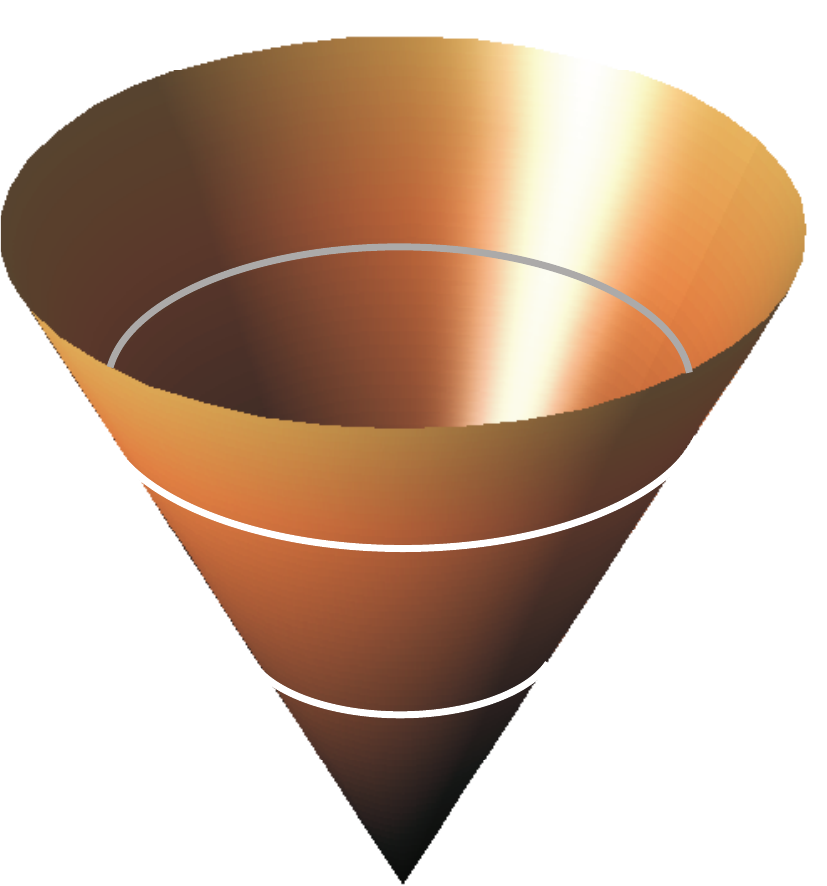}}
    \caption{The construction of $\frak{su}(2)$--invariant phase
      operators for irreps of the $(\lambda,0)$ type: equilateral
      triangles are turned into cones. The ``rings'' are made from
      weights in $\frak{su}(2)$--invariant subspaces.}
    \label{cone}
  \end{figure}

  One could also search for a complementary-based solution. Now, the
  $(2,0)$ irrep of $\mathfrak{su}(3)$ decomposes into a sum of two
  three-dimensional irreps of $\wp_{3}$, but the resulting matrices
  are incompatible with a polar decomposition of $\op{C}_{12}.$
  Indeed, one shows that the most general solution to the polar
  decompositions compatible with equation~(\ref{su320}) cannot produce
  commuting matrices.  This statement remains true for higher
  dimensional irreps of the type $(\lambda ,0)$, provided that
  $\lambda$ is finite.

  \subsubsection{Measuring non-commutativity.}

  Quite generally the phase operators do not commute. To quantify the
  amount by which, say, $\op{E}_{12} \op{E}_{31}$ fail to commute, we
  introduce the matrix norm $\Vert \op{M} \Vert^{2}= \Tr (
  \op{M}^{\dag} \op{M})$ and define
  \begin{equation}
    \op{M} = \op{E}_{12}  \op{E}_{31}  \, 
    \op{E}_{12}^{-1}  \op{E}_{31}^{-1} - \openone \,  . 
    \label{defineM}
  \end{equation}
  Obviously, $\op{M}$ should be the zero matrix if $\op{E}_{12} $ and
  $\op{E}_{{31}} $ commute.  To compare values of $\Vert \op{M}\Vert
  ^{2}$ for different irreps $(\lambda ,0)$, it is convenient to
  normalize the length of $\op{M}$ by dividing by the dimension
  $\frac{1}{2} ( \lambda +1 ) ( \lambda +2 ) $ of the irrep $(\lambda
  ,0).$ When this is done for the $\frak{su}(2)$-invariant solutions,
  we find
  \begin{equation}
    \frac{\Vert \op{M} \Vert ^{2}}{\frac{1}{2} ( \lambda +1 )
      ( \lambda +2 ) } = 2 \frac{ [ 2 ( \lambda +1) -1] }
    {\frac{1}{2}( \lambda +1 ) ( \lambda +2 ) } \, .
    \label{normalizednorm}
  \end{equation}
  This expression can be understood as follows. The action of
  $\op{E}_{12}$ commutes with the action of $\op{E}_{31}$ when the
  action of either is non--zero, as illustrated on the left of
  figure~\ref{su3su2phases}.  On the other hand, $\op{E}_{12}$ and
  $\op{E}_{31}$ do not commute if one or the other acts on some
  suitable ``edge'' state killed either by $\op{E}_{12}$ or by
  $\op{E}_{31}$. This is illustrated on the right of figure
  \ref{su3su2phases}.

  \begin{figure}
    \centerline{
      \includegraphics[scale=0.350]{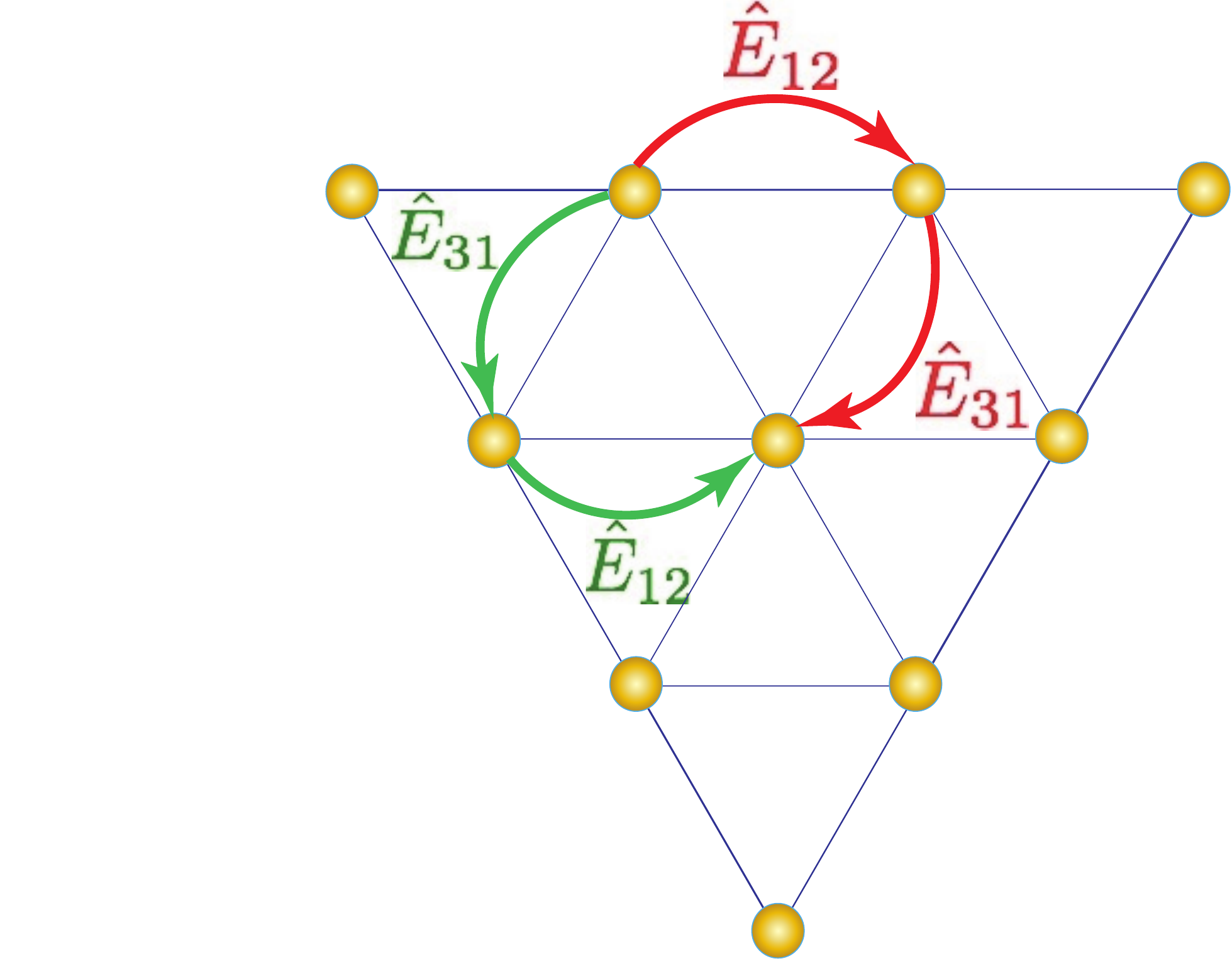}
      \includegraphics[scale=0.350]{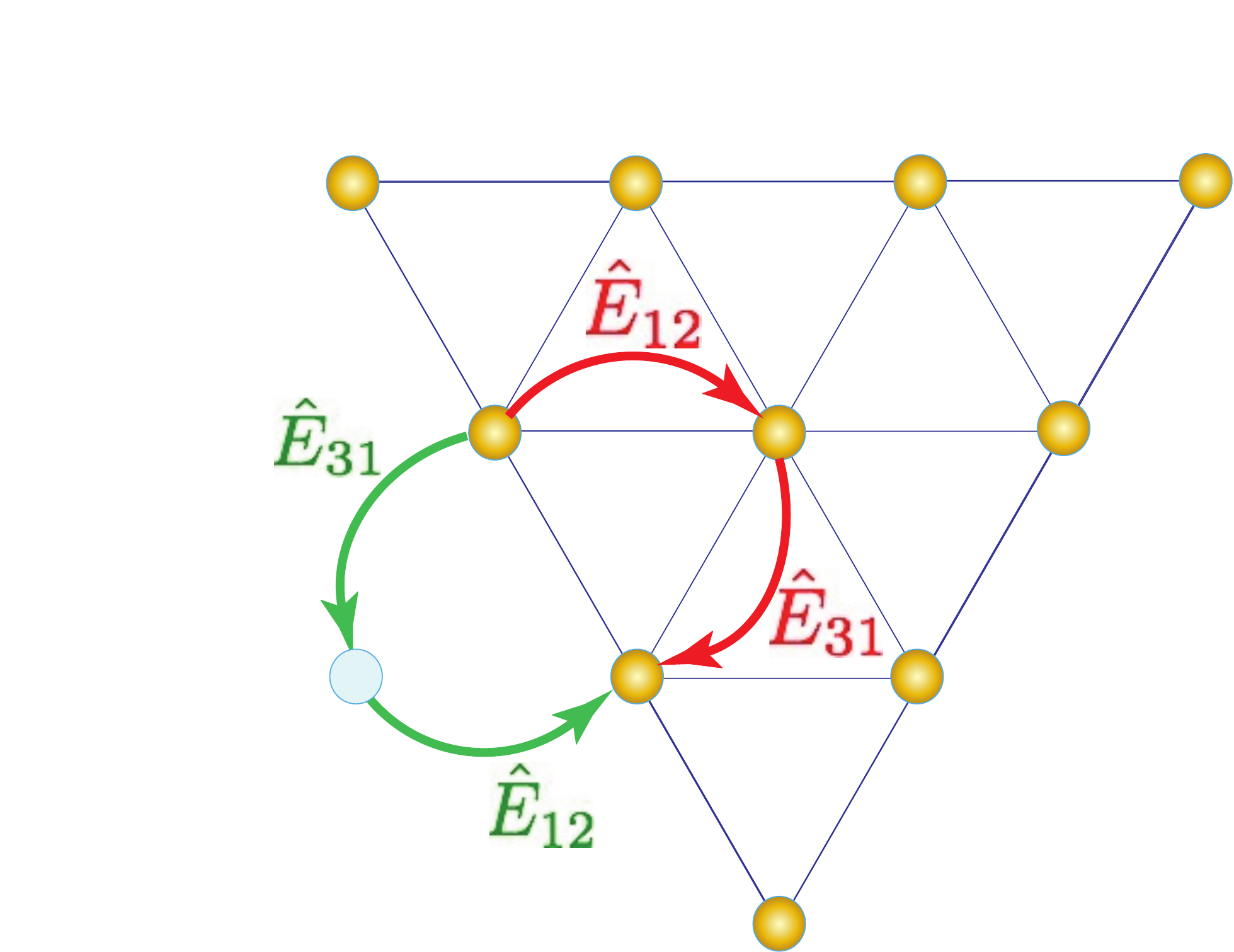}}
    \caption{Left: Illustrating how the operators $\op{E}_{12}$ and
      $\op{E}_{31}$ can commute. Red line: $\op{E}_{31} \op{E}_{12}$
      acting on a state.  Green line: $\op{E}_{12}\op{E}_{31}$. Since
      the matrix elements of $\op{E}_{ij}$ is always $1$, the
      operators commute in the case illustrated here.  Right:
      Illustrating how the operators $\op{E}_{12}$ and $\op{E}_{31}$
      can fail to commute when one or the other (or both) annihilate a
      state.  Here, the initial state is killed by $\op{E}_{31}$.}
    \label{su3su2phases}
  \end{figure}

  States killed by $\op E_{12}$ or $\op E_{31}$ are always located on
  the edge of the weight diagram for the irrep $(\lambda ,0).$ Any
  edge contains $( \lambda +1)$ states.  There are two problematic
  edges (one for $\op E_{12}$ and another for $\op E_{31}$) so the number of problematic states is $ 2 ( \lambda +1
  )$. Since these two edges have a single state in common, the number
  of problematic states, adjusted for double counting, is just $ 2 (
  \lambda +1 ) -1$. The overall multiplicative factor of $2$ comes
  from the calculation of the trace, and the denominator is clearly
  just the normalization factor.

  \subsubsection{The $\lambda \rightarrow \infty $ limit and its
    solution.}

  Equation~(\ref{normalizednorm}) shows that, for large $\lambda $,
  the amount by which $\op{E}_{12}$ and $\op{E}_{31}$ do not commute
  goes like $\sim \lambda ^{-1}$: for large $\lambda $, the
  $\frak{su}(2)$-invariant solutions commute and the phases
  $\varphi_{12}$ and $\varphi_{31}$ become additive. It is also clear
  that, as $ \lambda \rightarrow \infty $, the edge states become
  progressively displaced to infinity. In this limit, the finite
  triangular lattice of the weight diagram becomes a simple
  two-dimensional hexagonal crystal lattice.

  In the polar realization (\ref{su3polar}), the generators acting on
  states having finite weight simplify to
  \begin{eqnarray}
    \op{h}_{1} = - i \frac{\partial}{\partial \varphi_{1}} \, ,  
    \qquad
    \op{h}_{2} = - i \frac{\partial}{\partial \varphi_{2}} \, ,
    \nonumber \\
    \\
    \op{C}_{12} \sim \frac{1}{3} \lambda \e^{i(2\varphi_{1}-\varphi_{2})}
    \, , \quad
    \op{C}_{23} \sim \frac{1}{3}\lambda \e^{-i(\varphi_{1}-2\varphi_{2})} 
    \, , \quad
    \op{C}_{13} \sim \frac{1}{3}\lambda 
    \e^{i\left( \varphi_{1}+\varphi _{2}\right) } \, . \nonumber
  \end{eqnarray}
  In this limit, $\Gamma$ is Hermitian.  The rescaling $ \op{C}_{ij}
  \rightarrow \op{C}_{ij}/\lambda$ leads to commuting ladder
  operators.

  In the $\lambda \rightarrow \infty $ limit, the phase operators
  $\op{E}_{ij}$ act unitarily on every state of the form $
  \e^{i(n\varphi_{1}+m\varphi_{2})}$ with $n,m$ finite integers. The
  common eigenstates of $\op{E}_{12}$ and $\op{E}_{23}$ are
  \begin{equation}
    | \varphi_{1},\varphi_{2}\rangle =\frac{1}{2 \pi} 
    \sum_{n,m\in  \mathbb{Z}}
    \e^{i(2n-m)\varphi_{1}} \e^{i(2m-n)\varphi_{2}} \, . 
  \end{equation}

  \section{Concluding remarks}

  The general prescription provided for $\mathfrak{su}(3)$ can be
  extended to $\mathfrak{su}(n)$.  For instance, let us look briefly
  at $\mathfrak{su}(4)$.  The 12 roots corresponding to $\{
  \op{C}_{ij} : \, i \ne j= 1, \ldots, 4\}$ are located at the vertices
  of cuboctahedron, which is the intersection of a cube and an
  octahedron.  There are three diagonal operators, represented by
  three roots of length $0$ located at the center of the root diagram:
  $\op{h}_k=\op{C}_{kk}- \op{C}_{k+1,k+1}$.  Using again the boson
  realization $\op{C}_{ij}= \op{a}_i^\dagger \op{a}_j$ and boson
  states $| n_1 n_2 n_3 n_4 \rangle$, the diagonal operators
  correspond to population differences between consecutive levels.

  For such boson states, the weight diagram is a tetrahedron.  Each
  slice parallel to a fundamental root of $\mathfrak{su}(4)$ is an
  $\mathfrak{su}(3)$ subspace.  In particular, the action of some
  ladder operators will be undefined on one edge of the tetrahedron as
  the states on this edge are killed by them.  Two polar operators
  will thus fail to commute when they act on states in some specific
  edge of the weight diagram.

  There are $ (\lambda+1)(\lambda+2)(\lambda+3)/6$ boson states of the
  form $| n_1 n_2 n_3 n_4 \rangle$ with $n_1+n_2+n_3+n_4=\lambda$.
  There are $(\lambda+1)(\lambda+2)/2$ states on each edge of the
  weight diagram.  Two adjacent edges intersect on a line containing
  $\lambda+1$ states, and they have one point in common.  Defining
  $\op{M}$ as in (\ref{defineM}) for any pair of non-commuting roots
  and their phase operators, we found that, for $\frak{su}(4)$,
  \begin{equation}
    \frac{\Vert M\Vert ^{2}}{\frac{1}{6}
      (\lambda+1)(\lambda+2)(\lambda+3)}= 
    \frac{\left[ 2\times (\lambda+1)(\lambda+2) -
        \left( \lambda +1\right) -1\right] }{\frac{1}{6} 
      (\lambda+1)(\lambda+2)(\lambda+3)}.
    \label{normalizednormsu4}
  \end{equation}

  Our interpretation is thus that the non-commutativity of phases is
  an ``edge'' effect.  Since the number of points on an edge of the
  weight diagram grows with a rate $\sim\lambda^{n-2}$ while the
  number of states in an $\frak{su}(n)$ representation grows like
  $\sim\lambda^{n-1}$, phases operators constructed so as to preserve
  the $\frak{su}(n-1)$ subalgebras of $\frak{su}(n)$ will commute in
  the large $\lambda$ limit.

  As we have shown, in general, one cannot expect that phase operators
  will necessarily commute: they will commute in the limit of large
  representations where $\lambda\to\infty$, a limit which corresponds
  to a contraction of $\frak{su}(n)$ for which $\Gamma$ becomes
  Hermitian.

  We have not investigated in details the possibility of constructing
  commuting solutions which satisfy the complementary conditions for
  the special case of irreps of the type $(1,0,\ldots,0)$ of
  $\frak{su}(n)$.  However, it is probable that fully complementary
  solutions do not always exists.  To find phase operators that are
  pairwise complementary is closely related to the existence problem
  for mutually unbiased bases; likely, when $n$ is a prime, some
  elements of the generalized Pauli group are compatible with the
  polar decomposition of raising and lowering operators.  When $n$ is
  a power of a prime, it is not clear if the polar decomposition is
  compatible with the requirement of complementarity.  When $n$ is
  composite, the situation is even less clear as the construction of
  mutually unbiased bases remains an open problem.

  In conclusion, the coherent state representation used in this paper
  exhibits two nice features particularly relevant to the analysis of
  phase operators.  First, we have a geometrical interpretation of
  them in terms of azimuthal angles related to relative phases and
  associated with the ladder action of the appropriate generators.
  Second, the ``exponential part'' of the realization naturally
  provides the unitary part of the polar decomposition of the
  generators.  Note that the similarity transformation that maps the
  original $\Gamma$ realization into a Hermitian one simply rescales
  the diagonal entries of the polar part of the matrix representation
  of the generator and thus has no effect on the interpretation of
  phase part of the representation.
	
  The kind of coherent state representation used in this paper would
  seem to form a natural gateway into understanding phases in system
  described by algebras other than $\mathfrak{su}(n)$.  Certainly the
  geometrical structure of coherent states should allow one to
  interpret the  parameters of such a coherent state representation.
 
  The authors would like to thank Laura Toppozini and Alberto
  G. Barriuso for numerical work on some aspects of this problem.
  This work was supported in part by NSERC of Canada, the Grants
  FIS2008-04356 and FIS2011-26786 of the Spanish DGI and the UCM-BSCH
  program (Grant GR-920992).

  \appendix

  \section{Making the representation $\Gamma$
    Hermitian}\label{repGamma}

  In this appendix we provide some mathematical details on the
  representation $\Gamma$.  The discussion in this section is facilitated by denoting
  the representation of elements of $\mathfrak{su}(2)$ given in
  (\ref{eq:su2sub}) by $\gamma$.  In this notation,
  Eq.(\ref{Gammamap}) takes the form
  \begin{equation}
    \gamma(\op X)\vert \Psi\rangle\mapsto [\Gamma(\op
    X)\Psi]_{\vartheta}(\varphi)= 
    \langle\,\chi_j \vert \hat R_y(\vartheta)\hat R_z(\varphi)\gamma(\hat X)\vert \Psi\rangle
    \label{fullmap}
  \end{equation}
  The representations $\gamma$ and $\Gamma$ are clearly isomorphic.

  Since for $2j$ integer every representation of $\mathfrak{su}(2)$ is
  equivalent to a Hermitian representation, there must be an
  intertwining operator that takes the non-hermitian representation
  $\Gamma$ in (\ref{torus2}) to the hermitian $\gamma$.

  Following reference~\cite{Rowe:1989fr} we seek a similarity
  transformation
  \begin{equation}
    \gamma(\hat X)={\mathcal K}^{-1}\,\Gamma(\op X){\mathcal K}
  \end{equation}
  to bring $\Gamma$ to its Hermitian form.  Note that, under the inner
  product (\ref{iponthetorus}), $\Gamma(\op h)$ is already hermitian
  so ${\mathcal K}$ commutes with $\Gamma(\op X)$.  Next, from the
  requirement that $\gamma(e_+)=\gamma^\dagger (e_-)$, we find
  \begin{eqnarray}
    &{\mathcal K}^{\dagger}\Gamma^\dagger(\op e_-)({\mathcal
      K}^{-1})^\dagger&
    ={\mathcal K}^{-1}\Gamma(\op e_+){\mathcal K} \\
    \hbox{\rm or}\qquad 
    &{\mathcal K}{\mathcal K}^\dagger \Gamma^\dagger(\op e_-)
    &=\Gamma(\op e_+){\mathcal K}{\mathcal K}^\dagger\, .
  \end{eqnarray}
  Thus, the positive hermitian operator $S={\mathcal K}{\mathcal
    K}^\dagger$ is the intertwining operator for which
  \begin{equation}
    S\Gamma^\dagger(\op e_-)=\Gamma(\op e_+)S\, .
  \end{equation}

  Let $\bar \Gamma$ be the extension of $\Gamma$ to the infinite
  dimensional space spanned by the exponential functions:
  $\{\e^{im\varphi} : \, 2m \in \mathbb{Z} \}$.  On this space we also
    have
    \begin{equation}
      \bar S\Gamma^\dagger(\op e_-)=\Gamma(\op e_+)\bar S\, ,
    \end{equation}
    from which we obtain the recursion relation
    \begin{equation}
      \bar S_{m+1} (j+m+1)=\bar S_m\, (j-m)
    \end{equation}

    Because $\Gamma(\op e_-)\e^{-ij\varphi}=\Gamma(\op
    e_+)\e^{ij\varphi}=0$, it is clear that, if $\Psi(\varphi)$ is a
    function over the subspace of exponential functions spanned by
    $\{\e^{im\varphi}, -j\le m\le j\}$, then $\bar S\Psi(\varphi)$ is
    also in that subspace.  Thus, we can write $\bar S=S\cdot \Pi$,
    where $\Pi$ projects from the infinite-dimensional space of
    exponentials to the $\mathfrak{su}(2)$--invariant subspace
    $\{\e^{im\varphi}, -j\le m\le j\}$.  The intertwining operator $S$
    is then the positive part of $\bar S$, and hence invertible.  The
    similarity transformation ${\mathcal K}$ is the positive hermitian
    square root of $S$.  $\mathcal{K}$ is defined on
    $\{\e^{im\varphi}, -j\le m\le j\}$, and can be found recursively.

    One may verify that the diagonal matrix $\mathcal{K}$ with
    elements
    \begin{equation}
      \mathcal{K}_{mm} = 
      \displaystyle
      \sqrt{\frac{(2j)!}{(j+m)!(j-m)!}} \qquad \hbox{ if } -j\le m\le j\, ,\\
    \end{equation}
    satisfies the required recursion relation.

    By simple inspection one can backcheck the formalism and verify
    that this ``square root of a binomial'' matrix, which we write as
    \begin{equation}
      \mathcal{K} =\sum _{m=-j}^j
      \sqrt{\frac{(2j)!}{(j+m)!(j-m)!}}\, | jm \rangle\langle jm | \, ,
    \end{equation}
    satisfies $[ \mathcal{K}, \Gamma(\op{h}) ] = [ \mathcal{K} ,
    \Gamma(\op{e}_{-}) \Gamma( \op{e}_{+}) ] = [ \mathcal{K},
    \Gamma(\op{e}_+) \Gamma( \op{e}_-) ]=0$, and is such that
    \begin{eqnarray}
      \gamma (\hat h) \equiv  
      \mathcal{K}^{-1} \Gamma(\hat h) \mathcal{K} =  \Gamma (\hat h)   \, , 
      \nonumber \\
      \\
      \gamma (\hat e_-) \equiv  
      \mathcal{K}^{-1} \Gamma(\hat e_-) \mathcal{K} \, , 
      \qquad 
      \gamma (\hat e_+) \equiv  
      \mathcal{K}^{-1} \Gamma(\hat e_+) \mathcal{K} = 
      \gamma (\hat e_-)^\dagger \, . \nonumber
    \end{eqnarray}
    One immediately checks that the resulting $\gamma$ is indeed the
    standard Hermitian realization with action given in
    (\ref{eq:su2sub}).
  
    Having established that $\Gamma$ is equivalent to
    (\ref{eq:su2sub}), let us compare some aspects of their respective
    polar decomposition.  The matrix realization of $\Gamma
    (\op{e}_-)$ is factored as
    \begin{equation}
      \Gamma(\op{e}_-)= \Gamma (\op{E} ) \, \Gamma (\op{D})  \, , 
    \end{equation}
    where $\Gamma (\op{D}) = [\Gamma(\op{e}_-)^\dagger
    \,\Gamma(\op{e}_-)]^{1/2}$.  Following equations~(\ref{101}), we
    find
    \begin{eqnarray}
      \Gamma (\op{D})  &=&  \sum _{m=-j}^j (j-m)| jm \rangle\langle j m | \, ,  \\
      \nonumber  \\
      \Gamma( \op{E} ) &=&  \sum_{m=-j}^j  |j, m-1\rangle\langle j m | \,  ,\label{EGamma} 
    \end{eqnarray}
    In the same manner, using
    $\gamma(e_-)=\gamma(E)\sqrt{\gamma^\dagger (e_-)\gamma(e_-)}$, it
    is established that
    \begin{equation}
      \gamma ( \op{E} ) =  \sum_{m=-j}^{j} | j m - 1\rangle \langle j m | \, ,
      \label{Egamma}
    \end{equation}
    Equations~(\ref{Egamma}) and (\ref{EGamma}) are identical, but
    neither} is completely defined because the rank of the matrix is one
  less than the dimension $2j+1$ of $V_j$: the matrix element of $
  \op{E}$ calculated between the lowest weight state $| j j\rangle$
  and any other state $| jm\rangle$ is not determined.  We can make
  $\op{E}$ into a unitary matrix by taking $m$ modulo $2j+1$, and thus
  going from the finite line to the circle.  If we also impose the
  same cyclicity condition for $\Gamma(E)$, we find $\Gamma(\op E)$
  and $\gamma(\op E)$ coincide.  With the cyclic boundary conditions,
  the final result is:
  \begin{equation}
    \op E=\sum_{m} |j,m-1\rangle\langle j m| \, ,
    \quad -j\le m \le j, \  \hbox{\rm mod}(2j+1)\, .
  \end{equation}

  A direct extension of this argument applies to the Hermitian
  extension of the analogous representation (\ref{Gammasu3}) for
  $\mathfrak{su}(3)$. Details on the construction of the ${\cal K}$
  matrix for $\mathfrak{su}(3)$ can be found in
  reference~\cite{Guise:2002rt}.

  \newpage


\providecommand{\newblock}{}

\end{document}